\documentstyle[12pt,cite,doublespace,epsfig]{article}
\newcommand{\myslash}[1]{#1 \llap /}
\newcommand\partialslash{\partial \llap /}
\newcommand{\VEV}[1]{\left\langle #1\right\rangle}

\begin{document}

\begin{flushright}
       ADP-96-5/T210, hep-ph/9603260 \\
        {\em Talk given at the Joint Japan-Australia Workshop on 
             Quarks, Hadrons \\
             and Nuclei, Adelaide, South Australia, November 15-24, 1995\\
             (To appear in the conference proceedings)}
\end{flushright}

\begin{center}
{\large\bf Structure Functions of the Nucleon \\
in a Covariant Scalar Spectator Model} \\
\vspace{.5cm}
K.\ Kusaka$^a$, G.\ Piller$^{a,b}$, A.W.\ Thomas$^{a,c}$ and A.G.\ Williams$^{a,c}$\\
\vspace{.2cm}
{\it
$^a$Department of Physics and Mathematical Physics, University of Adelaide,\\
S.A. 5005, Australia
} \\
\vspace{.2cm}
{\it
$^b$
Physik Department, Technische Universit\"at M\"unchen, 
D-85747 Garching, Germany
} \\
\vspace{.2cm}
{\it
$^c$Institute for Theoretical Physics, University of Adelaide, 
S.A. 5005, Australia
} \\

\begin{abstract}

Nucleon structure functions, as measured in deep-inelastic lepton scattering, 
are studied within a covariant scalar diquark spectator model.  
Regarding the nucleon as an approximate two-body bound state 
of a quark and diquark, 
the Bethe-Salpeter equation (BSE) for the bound state vertex function  
is solved in the ladder approximation.  
The valence quark distribution is discussed in terms of 
the solutions of the BSE.

\end{abstract}

\end{center}
\vfill
\begin{flushleft}
\begin{tabbing}
E-mail: \={\it kkusaka, awilliam, athomas@physics.adelaide.edu.au}\\
\>{\it gpiller@physik.tu-muenchen.de}
\end{tabbing}
\end{flushleft}
\newpage

\section{Introduction}

In recent years many attempts have been made to understand 
nucleon structure functions as measured in lepton deep-inelastic 
scattering (DIS).  
Although perturbative QCD is successful in describing the 
dependence of structure functions on the squared momentum transfer, 
their magnitude is governed by the non-perturbative physics of 
composite particles, and is up to now not calculable directly from QCD. 

A variety of models have been invoked to describe nucleon structure functions.
The so called ``spectator model'' is a typical covariant approach 
amongst them\cite{spectatorModel}.  
In this approach the leading twist, non-singlet quark distributions are 
calculated from the process in which the target nucleon splits into 
a  valence quark, which is scattered by the virtual photon, 
and a  spectator system carrying baryon number $2/3$.  
Furthermore the spectrum of spectator states is assumed to be 
saturated through single scalar and vector diquarks.  Thus, 
the main ingredient of these models are covariant quark-diquark 
vertex functions.  

Until now vertex functions have been merely parameterized such that the 
measured quark distributions are reproduced, and no attempts have been made 
to connect them to some dynamical models of the nucleon.  
In this work we construct the vertex functions from a model Lagrangian 
by solving the Bethe--Salpeter equation (BSE).  However, we do not aim 
at a detailed, quantitative description of nucleon structure functions 
in the present work.  
Rather we outline how to extract quark-diquark vertex functions from 
Euclidean solutions of the BSE. 
In this context several simplifications are made.  
We consider only  scalar diquarks as spectators and restrict ourselves 
to the $SU(2)$ flavor group.  
The inclusion of vector diquarks and the generalization to 
$SU(3)$ flavor are relatively straightforward extensions 
and will be left for future work.  

The cross section for DIS of leptons from a nucleon is  characterized 
by the hadronic tensor $W^{\mu\nu}(P,q)$ 
where $P$ and $q$ are the four-momenta of the target and exchanged 
virtual photon respectively.
For unpolarized DIS, the hadronic tensor $W^{\mu\nu}$ is 
conventionally parameterized by two scalar functions 
$F_1$ and $F_2$.  
In the Bjorken limit ($-q^2, P\cdot q \rightarrow \infty$; 
but finite $x\equiv -q^2/(2\,P\cdot q)$) 
in which we work throughout, 
both structure functions depend (up to logarithmic corrections) on $x$ only,  
and are related via the Callan-Gross relation: $F_2 = 2 x F_1$.  

In the scalar diquark spectator model within $SU(2)$ flavor,   
the valence quark distributions 
are extracted from the hadronic tensor 
(Fig. \ref{fig:diquarkSpectatorForT}):
\begin{eqnarray}
        W^{\mu\nu}(q,P) & = &
                \left(\frac{5}{3}+\VEV{\frac{\tau_3}{2}}_N\right)\,
                \frac{1}{\pi}\,
                \int \frac{d^4k}{(2\pi)^4 i} \bar u(P,s) \bar\Gamma
                        S(k)\gamma^\mu \hbox{Im}\left(S(k+q)\right)
                        \gamma^\nu S(k)
        \label{compton}\\
        & & \qquad\qquad\qquad\qquad\qquad\qquad
                \times \hbox{Im}\left(D(P-k)\right) \Gamma u(P,s),
        \nonumber 
\end{eqnarray}
where the isospin matrix element $\tau_3$ has to be evaluated 
in the nucleon isospin space.  
We define  $u(P,s)$ as the target nucleon spinor and we use 
$S(k)=1/(m_q-\myslash{k})$ and $D(k)=1/(m_D^2-k^2)$ to denote 
the propagators of the quark and scalar diquark, 
respectively.  The integration runs over the quark momentum $k$, subject to 
on--mass--shell conditions for the diquark and the struck quark.  
Note that the vertex function $\Gamma$ and its PT conjugate $\bar\Gamma$ 
consist of two Lorentz scalar functions which depend on $k^2$ only because 
the diquark is on shell. 
In the next section we shall determine the vertex functions 
using a ladder Bethe-Salpeter equation.

\section{Scalar--Diquark Model for Nucleon}\label{diquarkModel}

We consider the following model lagrangian:
\begin{eqnarray}
        &{\cal L} = & \bar\psi_a\left(i \partialslash - m_q\right)\psi_a
                + \partial_\mu\phi_a^*\partial^\mu\phi_a - m_D^2 \phi_a^*\phi_a
        \label{action}\\
        & & + i \frac{g}{2\sqrt{2}} \epsilon^a_{bc}
                \psi^T_b C^{-1}\gamma_5 \tau_2\psi_c \, \phi^*_a
                -i \frac{g}{2\sqrt{2}} \epsilon^a_{bc}
                \bar\psi_b \gamma_5 C\tau_2\bar\psi^T_c \,\phi_a,\nonumber
\end{eqnarray}
where we have explicitly indicated color $SU(3)$ indices only.  
The symmetric generator $\tau_2$ of the flavor $SU(2)$ group acts on the 
iso--doublet field $\psi$ for the constituent quark 
carrying an invariant mass $m_q$.  
The charged scalar field $\phi$ denotes the flavor--singlet scalar diquark 
with invariant mass $m_D$.  

In this model the nucleon with the momentum $P$ and spin $s/2=\pm1/2$ 
is described by the BS vertex function $\Gamma$ (Fig. \ref{fig:BSvertex}):
\begin{equation}
        \delta_{ab} S(\eta_q P+q) D(\eta_D P-q) i \Gamma(q,P) u(P,s)
        =\int \frac{{\rm d}^4 x}{(2\pi)^{3/2}}e^{i k\cdot x}
                \left\langle 0\left| 
                        T \psi_a(\eta_q x)\phi_b(-\eta_D x)
                \right|P,s\right\rangle ,
\end{equation}
where we  set the weight factors to the classical values: 
$\eta_q = m_q/(m_q+m_D)$ and $\eta_D = m_D/(m_q+m_D)$.  
Then the vertex function $\Gamma$ obeys in the ladder approximation
the following BSE (Fig.\ref{fig:ladderBSE}):
\begin{equation}
        \Gamma(q,P)=g^2 \int\frac{{\rm d}^4k}{(2\pi)^4 i}
                S(-k-q-(\eta_q-\eta_D)P) S(\eta_q P+k) D(\eta_D P-k) 
                \Gamma(k,P).
        \label{ladderBSE}
\end{equation}
To solve the BSE for positive energy nucleon states we are free to choose the 
following Dirac matrix structure:  
\begin{equation}
        \Gamma(q,P)=\left[f_1(q,P)+
                \left(
                        -\frac{P\cdot q}{P^2}
                        +\frac{\myslash{q}}{\sqrt{P^2}}
                \right) f_2(q,P)\right]\Lambda^{(+)}(P),
        \label{define_f}
\end{equation}
with $\Lambda^+(P) = 1/2(1 + \myslash{P}/\sqrt{P^2})$, the projector 
onto positive energy nucleon states.  

We assume that the diquark and the nucleon are stable, namely 
$m_D < 2 m_q$ and $\sqrt{P^2} < m_q+m_D$.  
We then perform the Wick rotation of the relative energy variable and 
choose the nucleon rest frame: $P_\mu=P^{(0)}_\mu\equiv(P_0,\vec 0)$.  
The ``Euclidean'' functions $f_\alpha(\tilde q,P^{(0)})$ 
in terms of the momentum $\tilde q^\mu = (iq^4, \vec q)$ for a real 
$q^4 \in (-\infty,\infty)$ are then functions of 
$q_E \equiv \sqrt{-\tilde q^2} \in [0,\infty)$ and 
$z \equiv q_4/q_E \in [-1,1]$.  
We expand each of these functions ($\alpha=1,2$) as follows:
\begin{equation}
        f_\alpha(\tilde q, P) 
                = \sum_{n=0}^{\infty}\,i^n\,f^n_\alpha(q_E) 
                        C^1_n(z),
        \label{expanf}
\end{equation}
where $C^1_n(z)$ are  Gegenbauer polynomials and 
the phase $i^n$ is introduced for convenience 
since then the radial functions $f^n_\alpha(q_E)$ are real.  

The BSE in Eq.(\ref{ladderBSE}) then reduces to the following system of 
one--dimensional equations:
\begin{equation}
        \frac{1}{\lambda}\,f^n_\alpha(q_E)=
        \sum_{\beta=1}^{2}\sum_{m=0}^{\infty}\,
                \int_{0}^{\infty}{\rm d}k_E\, 
                {\cal K}^{n\,m}_{\alpha\,\beta}(q_E,k_E)\; f^m_\beta(k_E),
        \label{fn_eq}
\end{equation}
where we have introduced the ``eigenvalue'' $\lambda=(g/4\pi)^2$ 
for the quark--diquark coupling constant. The kernel function 
${\cal K}^{n\,m}_{\alpha\,\beta}(q_E,k_E)$ is a matrix whose 
elements are real and regular functions of $q_E$ and $k_E$.  By terminating 
the infinite series in Eq.(\ref{expanf}) at sufficiently high order, 
we can easily solve Eq.(\ref{fn_eq}) 
numerically as an ``eigenvalue'' problem for a fixed 
bound state mass, $P^2$.  To compare the magnitude of the radial functions, 
let us introduce 
the normalized $O(4)$ radial functions $F_n(q_E)$ and $G_n(q_E)$ together 
with the $O(4)$ spherical spinor harmonics 
${\cal Z}_{n\,j\,l\,s}$\cite{Rothe}.  
In the conventional gamma matrix representation we can write the nucleon 
solution at rest as: 
\begin{equation}
        \Gamma(\tilde q,P^{(0)})\,u(P^{(0)},s)
        =\sqrt{2}\,\pi
        \left(\matrix{\sum_{n=0}^{\infty}i^n\,F_n(q_E)\,
                                {\cal Z}_{n\,\frac{1}{2}\,0\,s}(\hat q)\cr
                \sum_{n=1}^{\infty}i^{n-1}\,G_n(q_E)\,
                        {\cal Z}_{n\,\frac{1}{2}\,1\,s}(\hat q) \cr}
                 \right),
        \label{O4diracrep}
\end{equation}
where $\hat q$ denotes angles for $\tilde q$ vector 
in the four--dimentional polar coordinate system.  
The factor $\sqrt{2}\,\pi$ is introduced such that  $F_n(q_E) = f_1^n(q_E)$,
while $G_n(q_E)$ is expressed as a linear combination of $f_2^{n-1}(q_E)$ 
and $f_2^{n+1}(q_E)$.  
Thus $f_1(q,P^{(0)})$ and $f_2(q,P^{(0)})$ correspond to the ``upper'' and 
``lower'' components of the nucleon Dirac field, respectively.

Now, let us consider the analytic continuation of $f_\alpha(q,P^{(0)})$.  
Since we are interested in applying the BS vertex function 
to the DIS process, we need to rotate the relative energy variable 
from on the imaginary axis back to on the real one.  
Recall that the sum over $n$ in Eq.(\ref{expanf}) converges even for a 
complex $z$, if $|z|<1$.  We can then analytically continue 
the Gegenbauer polynomials rewriting the argument $z\rightarrow \zeta\equiv
P\cdot q/\sqrt{q^2\,P^2}$ 
for the momenta satisfying $|\zeta| < 1$. 
For the radial functions we introduce new functions 
$\tilde F_n(q_E^2)\equiv F_n(q_E)/\,q_E^n$ and 
$\tilde G_n(q_E^2)\equiv G_n(q_E)/\,q_E^n$
\footnote{The fact that $\tilde F_n$ and $\tilde G_n$ are functions of 
$q_E^2$ was confirmed numerically.  }.
We analytically continue these functions by changing the argument 
$q_E^2\rightarrow -q^2$.  
We obtain the physical scalar functions $f_\alpha(q^2, P\cdot q)$ :
\begin{eqnarray}
        &  & f_1(q^2, P\cdot q)=\sum_{n=0}^{\infty}
                \frac{\tilde F_n(-q^2)}{(P^2)^{n/2}}\,
                        \left(\sqrt{q^2\,P^2}\right)^n\,C^1_n(\zeta),
        \label{f1Minkowski} \\
        &  & f_2(q^2, P\cdot q)=-\sum_{n=1}^{\infty}
                \frac{2P^2}{\sqrt{n(n+2)}} 
                \frac{\tilde G_n(-q^2)}{(P^2)^{n/2}}\,
                        \left(\sqrt{q^2\,P^2}\right)^{n-1}\,C^{2}_{n-1}(\zeta).
        \label{f2Minkowski}
\end{eqnarray}
Note that the Gegenbauer Polynomials together with the square root factors 
are polynomials of $q^2$, $P^2$, and $P\cdot q$, so that each term in 
the series (\ref{f1Minkowski}) and (\ref{f2Minkowski}) is regular and real 
as far as $\tilde F_n(-q^2)$ and $\tilde G_n(-q^2)$ are regular.  
We may then impose the on--mass--shell condition for the diquark and 
evaluate the sum over $n$. The resulting $f_1$ and $f_2$ are then functions 
of the squared quark momentum $k^2$ and can be  applied to the DIS process.  

However, the expressions (\ref{f1Minkowski}) and (\ref{f2Minkowski}) are 
valid only for momenta satisfying $ P^2 \;|q^2| > (P\cdot q)^2$.  
Also we found that a naive numerical sum over $n$, based on 
Eqs.(\ref{f1Minkowski}) and (\ref{f2Minkowski}) does not converge.  
Nevertheless, it can be shown that the vertex function $\Gamma$ for 
any kinematically allowed $k^2$ is regular 
when the diquark is on--mass--shell.  
This suggests that one may be able to continue some appropriate linear 
combinations of $f_1$ and $f_2$ outside of this kinematical range.  
Indeed, we found such a combination which we denote by 
$f_\alpha^{\rm on}(k^2)$.  
In terms of this on--shell scalar functions and the quark momentum, $k$, 
the vertex function $\Gamma$ together with the diquark on--mass--shell 
condition is given by
\begin{equation}
        \left. \Gamma\right|_{(P-k)^2=m_D^2} = 
                \left( f^{\rm on}_1(k^2) - 
                        \frac{2\myslash{k}}{\sqrt{P^2}}
                                f^{\rm on}_2(k^2) 
                \right)\Lambda^+(P).
        \label{DISpara}
\end{equation}
With this on--shell vertex function the valence contribution to  
the structure function $F_1(x)$ can now be calculated 
from Eq.(\ref{compton}).

\section{Numerical Results}\label{Results}

In this section we present our numerical results.  
For simplicity we considered an equal mass system; $m_q=m_D=m$, and we 
shall use $m$ as a unit for dimensionful quantities.  
To solve the BSE we used a $u$--channel form factor.  We replaced 
the quark--diquark coupling constant such that  
$g^2 \rightarrow g^2 \Lambda^2/(\Lambda^2 - u)$ with $\Lambda=2\,m$ and 
$u$ is the usual Mandelstam variable.  This form factor weakens 
the short range interaction between the constituents and 
ensures the existence of a discrete bound state spectrum 
for a large range of $P^2$.  

We solved Eq.(\ref{fn_eq}) as follows.  
First we terminated the infinite series in Eq.(\ref{expanf}) 
at some fixed value, $n_{\hbox{\small max}}$.  
Next we discretized the Euclidean momentum $q_E$ and $k_E$ and 
performed the integration over $k_E$ numerically together with some initially 
assumed radial functions $f^n_\alpha(k_E)$.  
This integral generated 
new radial functions and an ``eigenvalue'' $\lambda$ associated with them.  
We then used these functions as an input and repeated the above  
procedure until the radial functions and $\lambda$ converged.  

In Fig. \ref{fig:FandG} we plot the normalized $O(4)$ radial functions, 
$F_n(q_E)$ and $G_n(q_E)$, for the bound state mass 
$\sqrt{P^2} = 1.8 \,m$ as functions of $q_E$.  
It is clear that the magnitude of the radial functions with higher 
$O(4)$ angular momentum are  strongly suppressed 
compared with the lowest ones.  This justifies the truncation of 
the series in Eq.(\ref{expanf}).  It also confirms that 
the contribution to the ``eigenvalue'' $\lambda$ from the $O(4)$ radial 
functions with $n > 4$ is less than 1 \%.  The dominance of 
the lowest $O(4)$ radial function has been also observed 
in the scalar--scalar ladder model\cite{L+M}.  

In Fig. \ref{fig:fon} we plot the physical, on--shell scalar functions 
$f^{\rm on}_\alpha(k^2)$ for the bound state mass $\sqrt{P^2} = 1.8 \,m$ 
as functions of the squared quark momentum, $k^2$.  
These functions are calculated with the maximum $O(4)$ angular momentum 
$n_{\rm max}=4$.  
We found that the magnitude of $f^{\rm on}_1(k^2)$ and $f^{\rm on}_2(k^2)$ 
are almost the same even for weakly bound states.  
This result suggests that so--called ``non--relativistic'' approximations, 
in which one neglects the non--leading components of the vertex function 
($f_2$ in our model) are  valid only for extremely weakly bound states: 
$\sqrt{P^2}\sim 2\,m$. 
We have also confirmed that 
for weakly bound states ($\sqrt{P^2} > 1.8 \,m$) 
the dependence of $f^{\rm on}_\alpha(k^2)$ 
on $n_{\rm max}$ is negligible for a small spacelike $k^2$, 
e.g., $-k^2 < 5\,\,m^2$.
However, for large spacelike $k^2$, the convergence of the sum over $n$ 
becomes slow for any value of $P^2$ and numerical results 
for fixed $n_{\rm max}$ become less accurate.    
 
In Fig.\ref{fig:struct} we plot the valence quark distribution $F_1(x)$ 
for a  weakly ($\sqrt{P^2} = 1.8 \,m$, solid line)  
and for a  strongly bound state ($\sqrt{P^2} = 1.2 \,m$, dashed line).  
We have used $n_{\rm max}=4$ and the distributions are normalized such that 
the area below the curve is unity.  
For the weakly bound system, the valence quark distribution has a peak
\footnote{This peak will shift to $x\sim1/3$, if quark and diquark 
masses such as $m_D \sim 2\, m_q$ are used.  }
around $x\sim 1/2$.  
On the other hand, 
the distribution becomes flat for the strongly bound system.  
This behavior however turns out to be mainly of kinematic origin,   
since the distribution function is given by an  integral over the 
squared momentum $k^2$ carried by the struck quark with integration 
bounds  from $-\infty$ to $k^2_{\rm max}$, where 
\begin{equation}
        k^2_{\rm max}=x\left( P^2 -\frac{m_D^2}{1-x}\right).
\end{equation}
This kinematical bound determines the global shape of $F_1(x)$ to a large 
extent.

\section{Summary}

We have solved a BSE for a nucleon, described as a bound state of 
a quark and diquark within a covariant quark--scalar-diquark model.  
We have extracted the physical quark-diquark vertex function 
when the diquark is on--mass--shell from the Euclidean solution.  
The vertex function obtained was applied to a diquark spectator model for 
DIS, and the valence quark contribution to the structure function 
$F_1(x)$ was calculated.  
We found that the shape of the unpolarized valence quark distribution 
is mainly determined by relativistic kinematics and 
is independent of the detailed structure of the vertex function.

\noindent
{\Large\bf Acknowledgements}

This work was supported in part by the Australian Research Council.

\newpage
\noindent
{\Large\bf Figure Captions}

\begin{figure}[t]
	\centering{\epsfig{figure=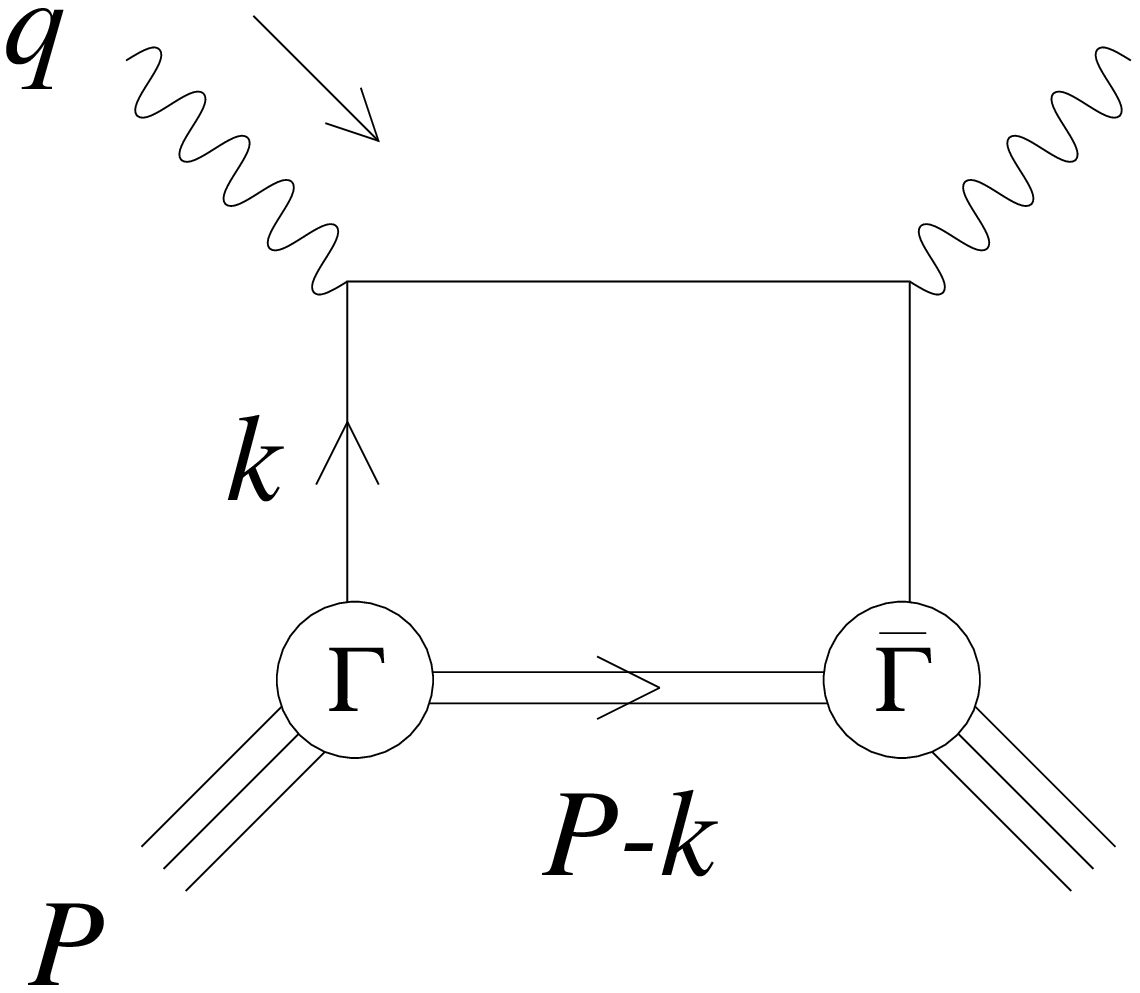,height=10.5cm} }
	\vspace{1.0cm}
	\caption{}
	\label{fig:diquarkSpectatorForT}
\end{figure}

\noindent
Fig. \ref{fig:diquarkSpectatorForT}:  
Diquark spectator process.

\begin{figure}[t]
	\centering{\epsfig{figure=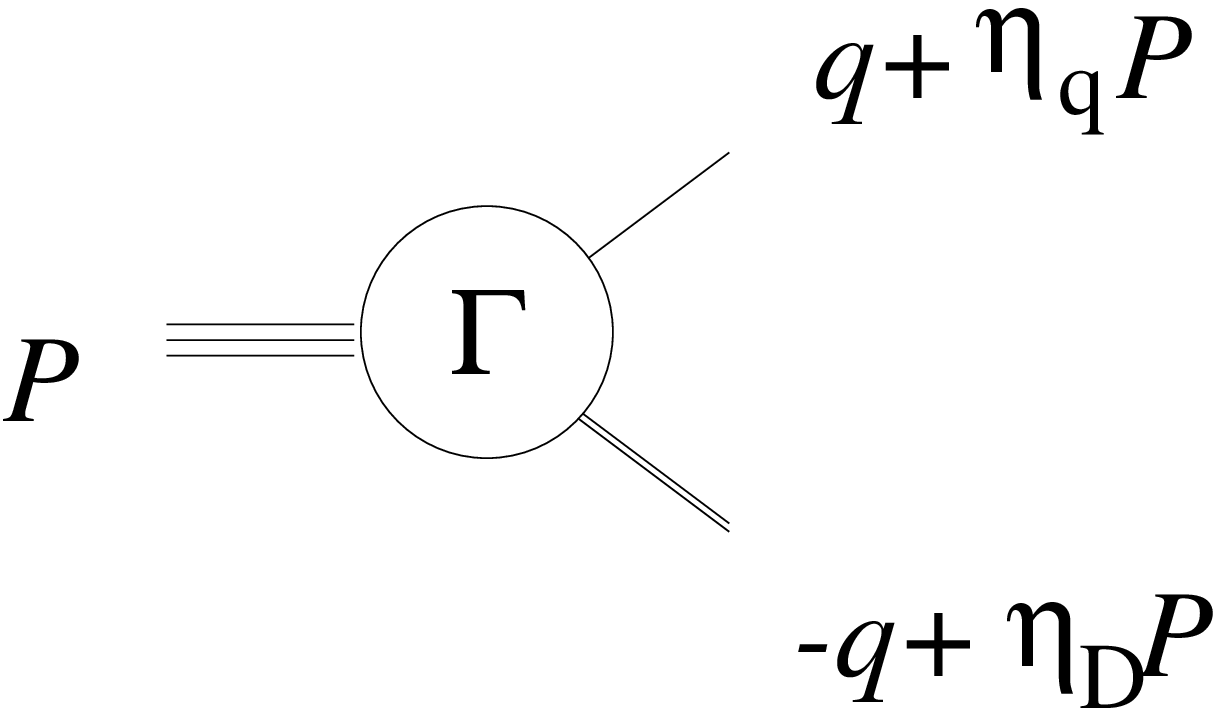,height=5.5cm} }
	\vspace{1.0cm}
	\caption{}
	\label{fig:BSvertex}
\end{figure}

\noindent
Fig. \ref{fig:BSvertex}:  
The Bethe-Salpeter vertex function.

\begin{figure}[b]
	\centering{\epsfig{figure=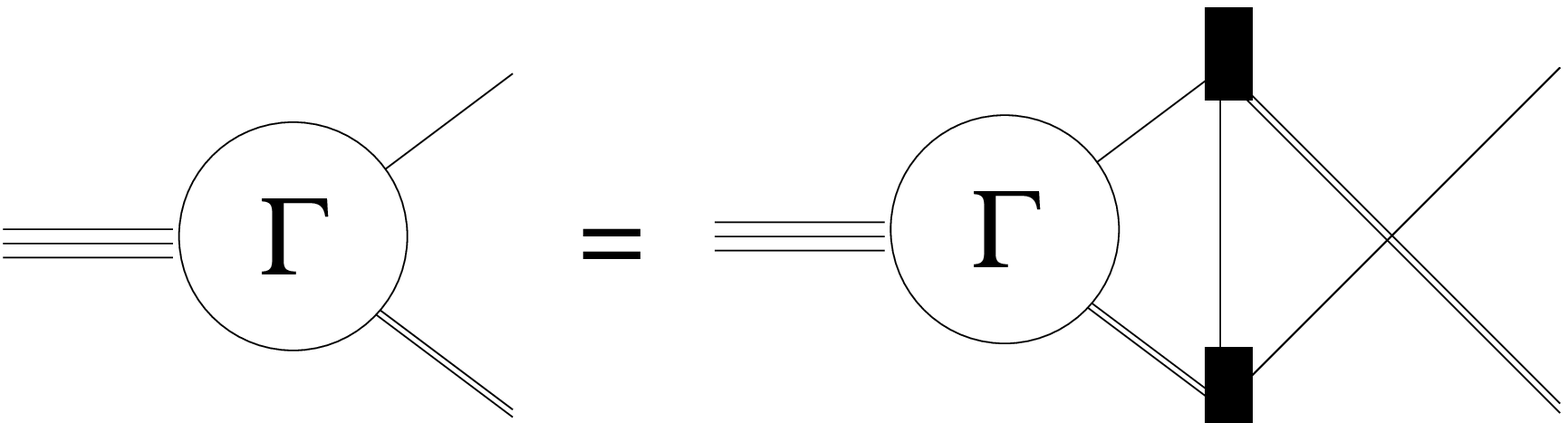,height=3.5cm} }
	\vspace{1.0cm}
	\caption{}
	\label{fig:ladderBSE}
\end{figure}

\noindent
Fig. \ref{fig:ladderBSE}:  
The Bethe-Salpeter equation in the ladder approximation.

\begin{figure}[t]
	\centering{\epsfig{figure=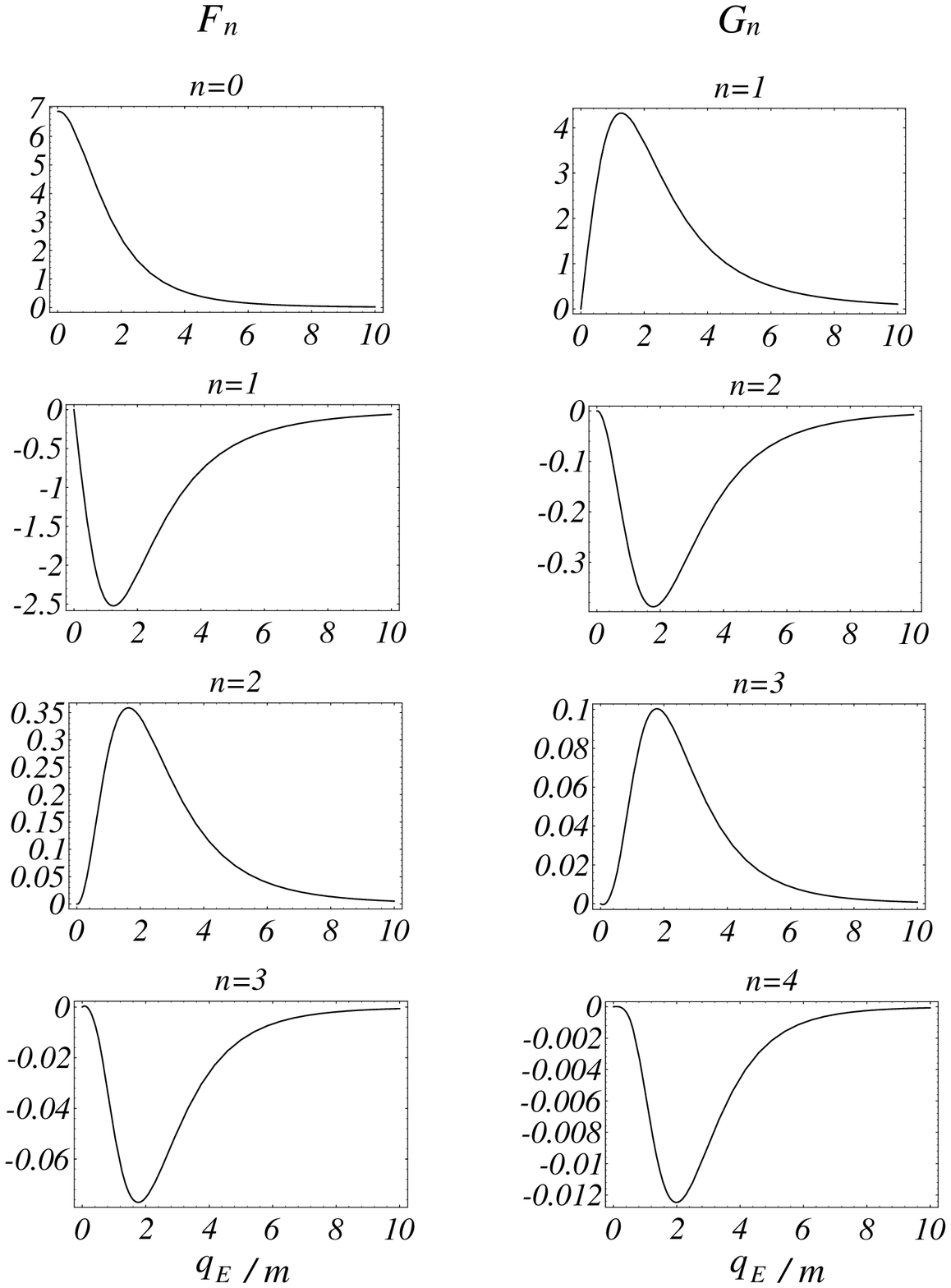,height=17.5cm} }
	\vspace{1.0cm}
	\caption{}
	\label{fig:FandG}
\end{figure}

\noindent
Fig. \ref{fig:FandG}:  
The normalized $O(4)$ radial functions $F_n$ and $G_n$ 
for the bound state with the mass $\sqrt{P^2}=1.8\,m$ as 
functions of the Euclidean momentum $q_E$.

\begin{figure}[t]
	\centering{\epsfig{figure=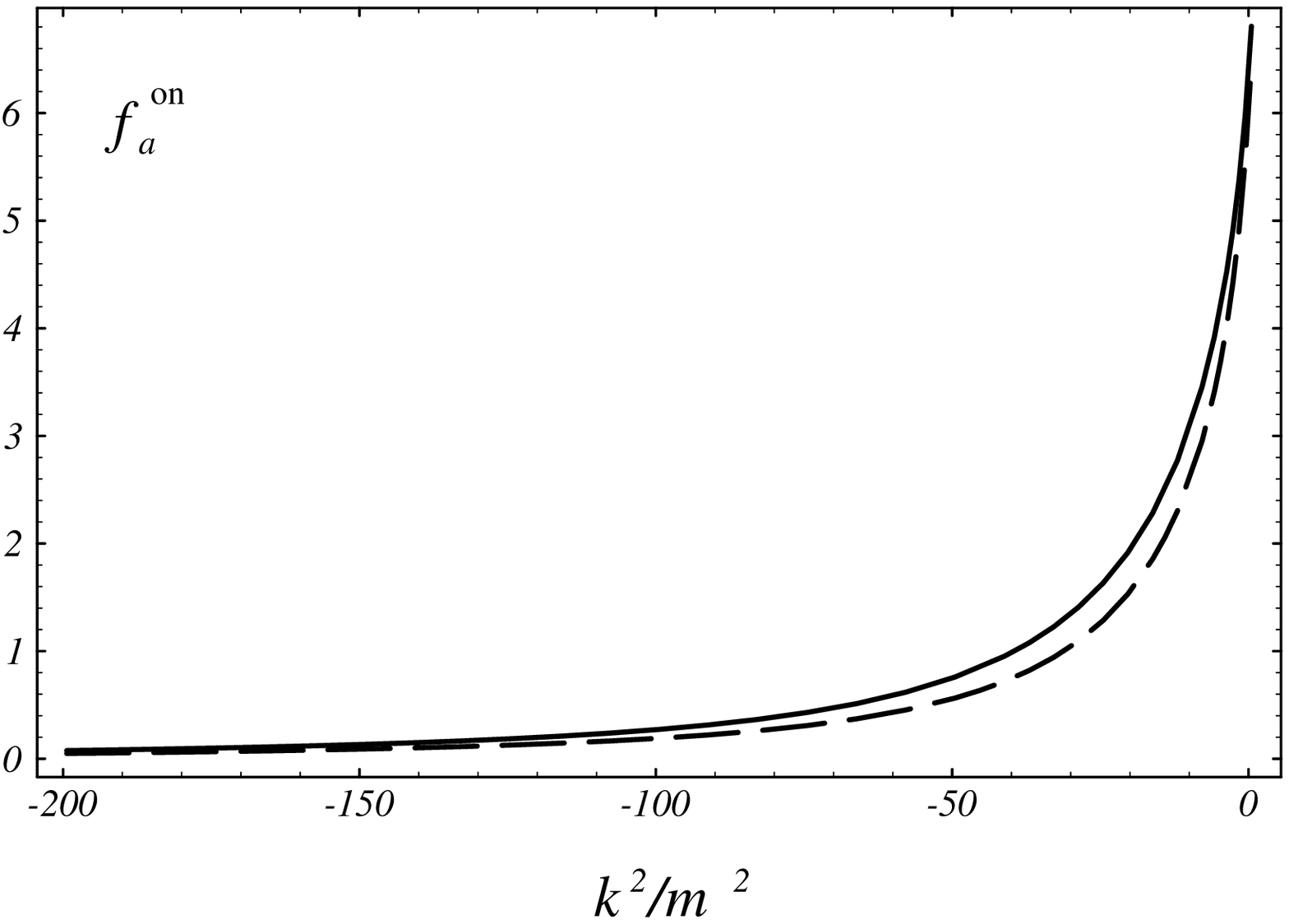,height=17.5cm} }
	\vspace{1.0cm}
	\caption{}
	\label{fig:fon}
\end{figure}

\noindent
Fig. \ref{fig:fon}:  
The on--shell scalar functions $f^{\rm on}_1(k^2)$ (solid) and 
$f^{\rm on}_2(k^2)$ (dashed) as functions of the quark momentum $k^2$.  
The mass of the bound sate is $\sqrt{P^2}=1.8\,m$ and the $O(4)$ angular 
momentum is truncated at $n_{\rm max}=4$.

\begin{figure}[t]
	\centering{\epsfig{figure=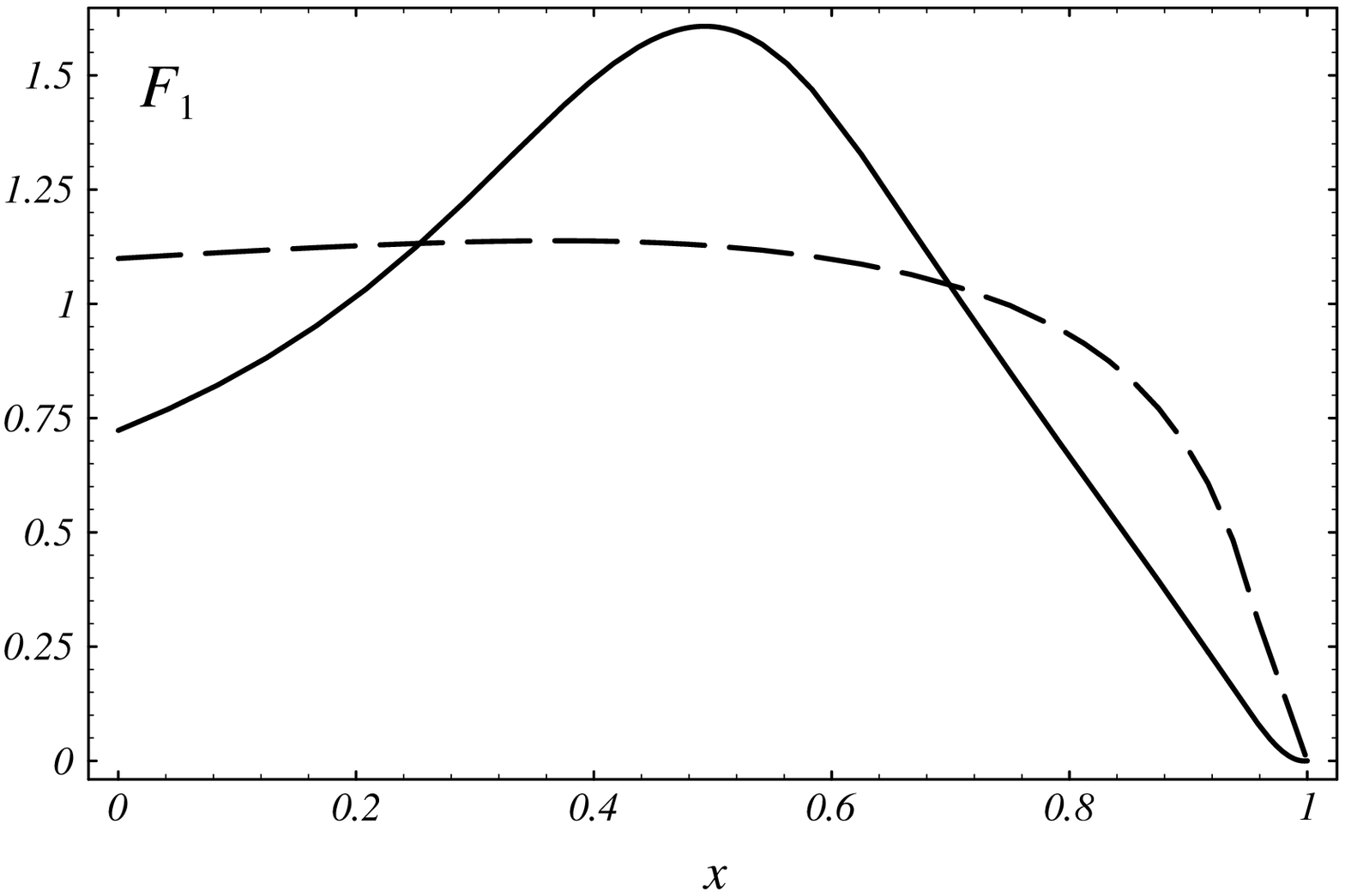,height=17.5cm} }
	\vspace{1.0cm}
	\caption{}
	\label{fig:struct}
\end{figure}

\noindent
Fig. \ref{fig:struct}:  
The valence quark contributions to the structure function $F_1(x)$.  
The solid (dashed) line denotes the weakly (strongly) 
bound state with the mass: $\sqrt{P^2}=1.8\,m$ ($1.2\,m$).

\newpage

\newpage

\newpage

\newpage

\newpage

\end{document}